\documentclass{article}
\usepackage{sao2,psfig}
\begin{document}
\setcounter{page}{39}
\issue{2000, 50, 39--50}
\markboth{Tikhonov, Makarov, Kopylov}{Study of clustering of galaxies, clusters and superclusters}
\title{Investigation of clustering of galaxies, clusters and superclusters
by the method of correlation Gamma-function}
\author{A.V. Tikhonov\inst{a} \and  D.I. Makarov\inst{b} \and A.I. Kopylov\inst{b}}
\institute{St.Petersburg State University \and \saoname}
\date{July 24, 2000}{October 30, 2000}
\maketitle
\begin{abstract}

Using the apparatus of correlation Gamma-function (``conditional density''),
we have analyzed spatial clustering of objects from several different
samples of galaxies, clusters and superclusters. On small scales the distribution
of objects obeys a power law drop of density with a power index of 0.9--1.5.
On a scale of $\sim 30$\,Mpc for independent samples of bright galaxies
and clusters of galaxies we have
detected a pronounced break in the slope of the Gamma-function with a power
index decreasing to $\sim 0.3$. The clustering is much less pronounced
in the region from 40 to 100\,Mpc, and  there is reason to suppose that the
distribution of objects changes to homogeneous on scales larger than 100\,Mpc. This is indicated
by the slope of the Gamma-function close to 0 for a sample of rich clusters
of galaxies up to 250\,Mpc.

The slope of the Gamma-function prior to the break, which characterizes
the degree of clustering of matter, changes essentially and in a complex
manner when passing to brighter (massive) objects. This suggests that the
large-scale structure of the visible Universe even on small scales is considerably
more complex than the fractal distribution described by one dimension
(monofractal).
\end{abstract}
\section{Introduction}
One of the principles on which standard cosmological models are based is
homogeneity of distribution of matter on large scales (Peebles, 1983).
A direct indication of such a character of distribution of matter in the observed
Universe is, for instance, the homogeneity and anisotropy of CMB radiation.
At the same time, the investigators of the large-scale structure have discovered
inhomogeneities in the distribution of visible matter on scales to 100\,Mpc
(IAU Symp., 1978; Geller \& Huchra, 1989).
Within the frames of the present-day knowledge there are two main large-scale
structure models: a homogeneous model with fluctuations of density of finite
amplitudes and a fractal model characterized by self-similarity of structures of the
observed Universe in a certain interval of scales.
A question on scales of extension of fractal structures is being debated.
According to Davis (1997),
the self-similarity properties of distribution of galaxies manifest themselves
but to 10\,Mpc. Pietronero et al. (1997) advocated the idea that a maximum
scale of clustering cannot be reached by the present-day surveys of galaxies
and clusters of galaxies, the fractal properties in the distribution of
matter are observed to $\sim 150$\,Mpc and there are signs of extension
of the fractal structure to $\sim 1000$\,Mpc.

There are several key stages in the history of origin and development of
knowledge of fractal (self-similar) type of distribution of matter. Having
noticed the sequence of clustering (galaxies, groups, clusters, superclusters),
de Vaucouleurs (1970) provided an observational substantiation of the model
of hierarchical clustering which could explain the power, with a slope
$\gamma \sim 1.7$, decreasing of density with increasing radius (de Vaucouleurs
diagram). By 1982 Mandelbrot (1977, 1982) had extended these ideas and introduced
a mathematically rigorous notion of fractals and proposed, following from
general considerations, a fractal dimension $D=1$ for the Universe.

About 20-year work over the redshift surveys, increasingly deeper and
complete, made it possible to see in the distribution of galaxies the structures
and voids of different shape (Huchra et al., 1983), the filling of the
volume by objects with measured $z$  still more clearly defined the features
noticed previously (Fairall, 1998). The maps of the distribution of galaxies
in space of redshifts (CfA1 survey (Huchra et al., 1983), the first deep
two-dimensional (slice) CfA2 survey (Geller \& Huchra, 1989) and others)
were consistent with intuitive ideas of stochastic fractal point distributions
in three-dimensional space: extended ``coherent regions'' of increased density,
voids occupying a considerable part of the volume, general ``clumpy''
and irregular pattern of distribution.

The more detailed construction of the three-dimensional pattern of the
nearest regions of the Universe gave impetus to development of a variety
of statistical methods (Borgani, 1995; El-Ad et al., 1996; Paladin \&
Vulpiani, 1987; Plionis \& Valdarnini, 1992) of isolation and study of
different structures in the distribution of objects in the Universe
(such as voids, filaments, superclusters, ``walls''). As distinct from the
two-point correlation function, these methods reveal more finely
characteristic morphological properties of the structures. The methods of search for
fractal (scale-invariant) properties of matter distribution were executed
mathematically (Borgani, 1995; Pietronero, 1987; Coleman \& Pietronero, 1992).

\section{Mathematical apparatus}
If the mean density in the volume under study is not determined, i.e. it
varies considerably with increasing working volume up to scales characterizing
the volume
under study, the standard statistical method of analysis of the large-scale
structure, the two-point correlation function ($\xi$-function) (Peebles,
1983; Davis \& Peebles, 1983; Boerner \& Mo, 1990; Klypin \& Kopylov, 1983;
Dalton et al., 1992) will then yield a result dependent on
the parameters of the sample and the way of calculation. For instance,
it has been noted that the ``correlation scale'' $r_0$ grows with increasing
volume of the sample (Einasto et al., 1986; Coleman \& Pietronero, 1992).

Pietronero et al. (1988) and Coleman \& Pietronero (1992) noticed that if
the given sample does not include the scales where the amplitude of clustering
is small, i.e. the distribution is homogeneous (the scatter of the number of
galaxies in distant equal volumes is described by the Poissonian statistics),
then the parameters of the $\xi$-function do not show the real amplitude and the
limiting scale of clustering, and proposed another method of calculation
of the correlation function of density, the so-called Gamma-function (hereafter
Gamma) or the ``conditional density'' (Coleman \& Pietronero, 1992) used in
statistical physics for analysis of non-analytical structures with large-scale
density correlations. Pietronero and his co-workers (Pietronero et al., 1997;
Sylos Labini et al., 1996, 1997; Montuori et al., 1997; Sylos Labini et al., 1998)
had been using this method for over 10 years to study the distribution of astrophysical
objects and concluded that fractal structures are extended to scales
 100--150\,$h^{-1}$\,Mpc, and from indirect data (relation between ``radial'' density
and distance from ESP survey) even to $\rm \sim 1000\,h^{-1}$\,Mpc (Pietronero
et al., 1997).

The papers of Pietronero and his colleagues found a broad response. Cosmological
models were proposed which took account of the fractality (Baryshev, 1981;
Baryshev et al., 1994) as an essential part of the Universe picture
(interpretation of redshift as a gravitation effect which is determined
by the global inhomogeneity of the Universe but not by the expansion of
space). Controversy developed as to the scales of extension of the fractal
structure (Davis, 1997; Pietronero et al., 1997). Nevertheless, among the
researchers of the large-scale structure only Pietronero's group keeps
holding ``radical'' viewpoints of the enormous extension (up to 1000\,Mpc)
of the fractal law of density variation with distance, which is described
by the single dimension $D$ (Mandelbrot, 1977; Feder, 1988).

The fractal
methods are difficult to apply to the study of distribution of galaxies and clusters of
galaxies because the existing samples represent the distribution of a
finite number of points in a finite volume, whereas the fractal properties
in terms of mathematics are determined at the limit of infinity. For the
``physical'' or ``dynamical'' fractal a large (but finite) interval of
scales between the lower and upper limits of manifestation of self-similarity
properties is necessary ( McCauley, 1997).

The condition ``volume limited'' should be met in order that all regions
of the sample be represented with equal rights, that is, from every object
one could ``see'' any other. For samples of galaxies this condition is
realized when the objects, whose absolute stellar magnitude is larger than that of the
faintest galaxies at the far limit of the sample, are rejected.

It is important to note that the ideology of construction of the Gamma
calls for abandoning a priori assumptions concerning the properties of the
distribution of the sample objects. Thus for the method to be used,
careful  preliminary work is required on the creation of a sample of
homogeneously selected objects with sharply defined spatial boundaries.
The scales to which the Gamma can be calculated are restricted by the radius of a
maximum sphere with the centre in the object of the sample, which can still
fall within the boundaries of the sample.

The Gamma is variation of the mean density of objects as the
volume under study increases. The differential and integral Gamma-functions
are determined as follows:
\begin{equation}
\Gamma(r)=\frac{1}{N}\sum\limits_{i=1}^{N}
  \frac{1}{4\pi r^2 \Delta}\int\limits_{r}^{r+\Delta}n(r_i-r)dr,
\end{equation}
\begin{equation}
\Gamma^*(r)=\frac{1}{N}\sum\limits_{i=1}^{N}
  \frac{1}{4\pi r^3}\int\limits_{0}^{r+\Delta}n(r_i-r)dr,
\end{equation}
where $n(r)=\frac{1}{N}\sum\limits_{i=1}^{N}\delta(r_i-r)$ is the numerical
density.

The integral function ($\Gamma^*$) averages the contribution of different scales
and therefore, with the presence of properties of a monofractal in the distribution,
it smoothes fluctuations and allows the dimension of the distribution
to be measured with a higher accuracy than the differential function. At the same
time
the integral function is not very sensitive
to change of the type of distribution. The differential function, in principle,
registers better change of the mode of distribution than the integral
function does, and in this sense it is more informative, but at the same
time it is more susceptible to fluctuations.

In some cases the form of the Gamma has a simple interpretation. For the
monofractal with the single dimension $D$ (``filling of the volume''
occurs, on the average, in the same manner on all scales for all objects) the
relationship between the number of objects in a sphere and its radius,
$N(r)=Br^D$, is satisfied. In the general case $D\le 3$. For a homogeneous
distribution the number of points in the volume is directly proportional
to the volume ($D=3$).

The average density in a sphere of radius $R_s$ placed in a given fractal
structure or, for instance, in a ~uniform ~~distribution~ ~is
$\langle n\rangle=\frac{N(R_s)}{V(R_s)}=\frac{3B}{4\pi}R_s^{-(3-D)}$.
At $D<3$ the average density $\langle n\rangle$ is a diminishing function
of radius $R_s$ and $\langle n\rangle \rightarrow0$ at $R_s\rightarrow\infty$
is satisfied for each point regarded as the centre of the sphere of radius
$R_s$. For the homogeneous distribution ($D=3$) the average density is constant
and independent of the volume it is measured in.

In the case of  a ``pure'' monofractal structure, $\Gamma^*$
must change according to a power law with the slope of the straight
line in the plot $\log(\Gamma^*)-\log(r)$, which is defined by the fractal
dimension of the set. If the fractal properties manifest themselves
only to a certain scale $\lambda_0$ with transition to uniformity on
larger scales, the relations:
$$\Gamma(r)=\frac{DB}{4\pi}r^{-(3-D)},r<\lambda_0,$$
$$\Gamma(r)=\frac{DB}{4\pi}\lambda_0^{-(3-D)},r\ge\lambda_0$$ are satisfied.
The Gamma function is represented here by definition in the form
$\Gamma(r)=\frac{1}{4\pi r^2}\frac{dN}{dr}$ and $N(r)=Br^D$.

When analysing the real distribution, it should be borne in mind that even
if the points of the integral Gamma fall well on a straight line in the
region of drop of the ``conditional density'', it does not suggest that
the properties of the monofractal are present in the distribution.
Analysis of artificial distributions shows that the power law Gamma diminishing
with the index $-\gamma$ is a necessary, but not a sufficient condition for
fractality of the distribution with the dimension $D=3-\gamma$.
Nevertheless, the power law character of the Gamma in the interval of scales,
where it is defined with sufficient reliability, admits a fractal interpretation.
Other attributes of fractality, for instance, self-similarity of the structures,
the proportion of voids in the volume, require a special study by suitable
methods.

The paper consideres the relations between
$\log(\Gamma)$ and $\log(r)$, $\log(\Gamma^*)$ and $\log(r)$.
The angular coefficient of the approximation line, which is plotted on the
chosen region of the $\log(r)$ variation, defines the correlation dimension
of the distribution (co-dimension). The greater slope (corresponding
to the smaller dimension implies, on the average, a stronger decrease of density
inside the volume and therefore a higher clustering of objects. The horizontal
regions of the plot point, in the general case, to uniformity of distribution
of objects in the sample on  corresponding scales.

Fig. 1 shows the expected behaviour of the Gamma in transition from fractal
clustering on scales less than $\lambda_0$ to uniform distribution on larger
scales. The figure illustrates the informativeness of using the Gamma in
searching for the scale of change of the condition of clustering $\lambda_0$,
provided that $R_s$ is well larger than $\lambda_0$.

\begin{figure}
\centerline{\psfig{figure=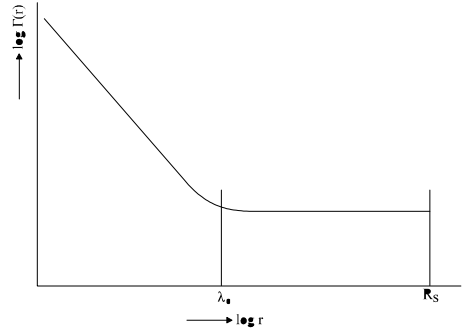,width=6cm}}
\caption{$\lambda_0$  is the maximum scale of extension of correlations.
	 $R_s$ is the limiting scale of calculation of gamma-functions.}
\label{Fig1}
\end{figure}

\section{The way of calculation}
\begin{itemize}
\item The boundaries of the region of the highest completeness are denoted (the zone of
incompleteness along the radial axis and the region of strong interstellar
extinction are usually excluded).
\item The density of the objects that fall within the spherical
layer $r_0 < r < r_0+\Delta$ around each object, that is, the density at a
given distance from the object of the sample for the differential Gamma,
and the density in spheres of radius $r_0$ with the centres in the objects
of the sample for the integral Gamma are calculated (the centres of the spheres are not included in the counts,
i.e. the density of the ``neighbours'' is measured). The calculations are averaged.
The results are presented on a logarithmic scale in the form of the relations
$\log(\Gamma)$ and $\log(r_0)$, $\log(\Gamma^*)$ and $\log(r_0)$.
\item If a part of some sphere falls outside the sample limits with increasing
working radius $r_0$, the sphere is then excluded from the calculations. Thus,
with increasing working radius, only the spheres with the centres progressively
nearing to that of the volume containing the sample are involved in the calculations.
The calculations are terminated when the number of remaining spheres is
$N_{sp} < 10$.
\end{itemize}

The method of calculation implemented in the present paper differs from the
variant used by Pietronero and his colleagues but in insignificant
details (the step of increasing the working radius, the number of finite
spheres at which the calculations are terminated and others).

We call this method of Gamma calculation ``classical'' as distinct from
modifications using various kinds of weighting of the regions of spheres
falling outside the sample limits (Lemson \& Sanders, 1991; Andreani et al.,
1991) and other procedures allowing more effective utilization of the whole
volume of the sample. The demerit of the method is that the sample objects
are unequal in rights since the objects located at the boundaries fall out
rapidly from the calculation as the centres of spheres. Therefore the result
on large scales is strongly dependent on the distribution of a small number
of objects near the centre of the volume being studied. From our
estimates the Gamma remains informative at a minimum number, 100--200,
(depending on the volume) of objects, left in the sample. A detailed examination
of the question is presented in Sylos Labini et al. (1996).

The problem of determining the error of approximation of the Gamma slope
in the region, where  the power law decrease of density is observed,
is difficult to overcome because there is no ``etalon'' of distribution for the Gamma,
as distinct from the statistics which compare in some way the distribution of a sample
with uniform one. The dispersion of the Gamma calculations with respect to the mean
does not, strictly speaking, indicate the error of the mean value of the
Gamma since for an arbitrary distribution, for instance, fractal, the spread
in values may be natural, and only the mean Gamma values are of significance.
The  ``bootstrap'' method (Ling et al., 1986; Mo et al., 1992) may possibly
give  correct estimates of the errors.

The distances for all samples of galaxies were taken Euclidean and were
determined by the Hubble law $R = V_0 / H_0$, where $V_0$ is the radial
velocity, and $H_0$ is the Hubble constant. The K-correction was not taken
into account.

When changing the redshift values to metric distances for the clusters
(Abell and APM), the following formula was used:
\begin{equation}
R=\frac{c}{H_0}\frac{q_0+(q_0-1)(\sqrt{1+2q_0z}-1)}{q_0^2(1+z)}.
\end{equation}

The values $H_0 =100$\,km/s/Mpc for the Hubble constant and $q_0 = 0.1$
for the deceleration parameter were used in the calculation by this formula.

\section{Description of samples and results of calculations}
To analyse the Gamma behaviour, diverse data were chosen which represented
samples of physically isolated structures of luminous matter (galaxies, clusters,
superclusters), allowing the clustering on scales from 50\,kpc to $\sim 200$\,Mpc
to be estimated.

\subsection{Local Volume}
This is a sample of nearby galaxies with radial velocities less than 500\,km/s
with respect to the Local Group centroid. The original list made by Kraan-Korteweg
and Tammann (1979) numbered 179 objects. Further it was substantially
complemented by Karachentsev (the most complete compilation can be found in Makarov's
thesis (2000). We will emphasize that this is the only sample with photometric estimates of distances
of galaxies. We used a sample of 330 objects. Of these 330 objects, 193
galaxies have photometric estimates of distances, for 47 objects the distances
are taken from their membership in known groups, and only for the rest
of the objects Hubble distances   ($H_0=70$\,km/s/Mpc) were used.

A supergalactic ``disk'', in which 80\,\% of galaxies are concentrated, is
dominated in the distribution of the Local Volume galaxies. This is a flat
structure that occupies the centre of the volume and becomes more dense
in the direction toward Virgo. Practically all known groups of galaxies of the
Local Volume (the Local Group, M81, NGC\,5128+NGC\,5236 (Centaurus),
NGC\,4244+NGC\,4736 (Canes Venatici), M101) are located in the disk. In
the northern direction (in supergalactic coordinates) the
``Local Void'' detected by Tully is situated. It takes approximately half of the volume
in question. From Karachentsev's estimates there are no galaxies brighter than
$M_B = -13^m$ in absolute magnitude in this region.

The results of the Gamma-analysis are presented in Figs. 2 and 3. The slope
$\gamma\sim1.2$ in the Local Volume is likely to be defined by the distribution
geometry. The expected slope for a pancake-type structure is $\gamma\sim1$.
The test with the use of random ``mixing'' of galaxies in the ``disk'' does
not change essentially the result of the Gamma-analysis. That is, strictly
speaking, the distribution of galaxies in the Local Volume is largely
inhomogeneous because of the presence of the ``Local Void'', but not fractality.
It is likely that the slope of the Gamma towards small scales remains practically up to
the size of a large galaxy ($\sim 50$\,kpc) owing to the dwarf companions of
the most massive galaxies.

\begin{figure}
\centerline{\psfig{figure=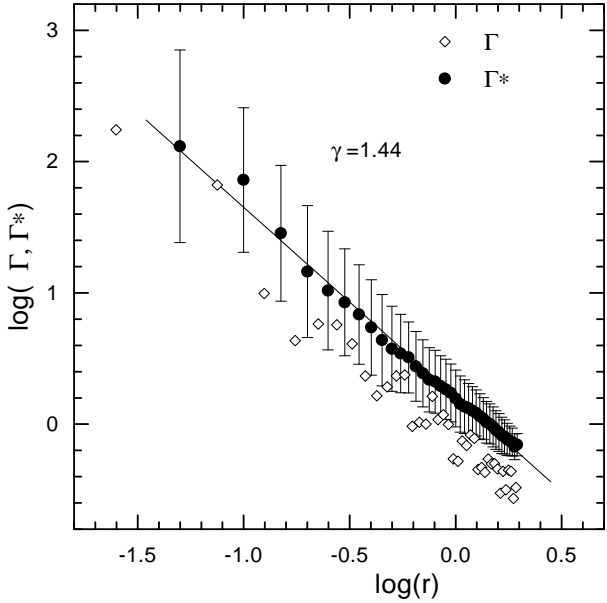,width=8cm}}
\caption{The ``Disk'' of the Local Volume $R_{\mbox{s}}=2.0$~Mpc.}
%\label{Fig2}
\end{figure}

\begin{figure}
\centerline{\psfig{figure=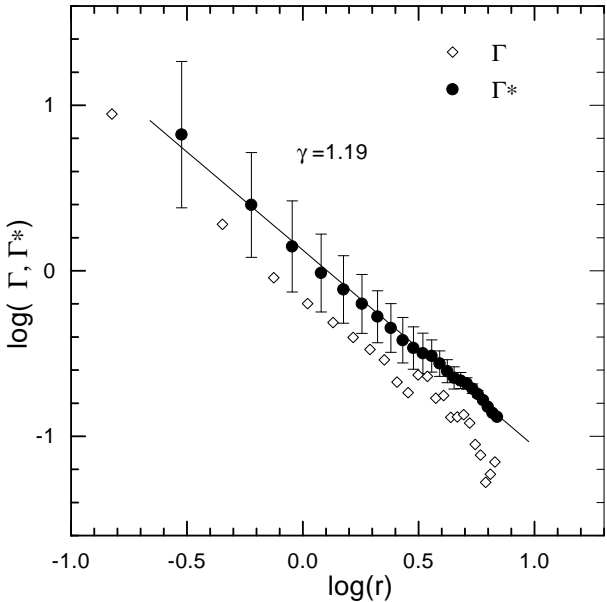,width=8cm}}
\caption{The Local Volume $R_{\mbox{s}}=7.15$~Mpc.}
%\label{Fig3}
\end{figure}

\subsection{CfA2 and SSRS2 catalogues of galaxies}
{\bf CfA2} (a subsample from the UZC catalogue (Falco et al., 1999)).
The catalogue of Zwicky with $m_{\rm Zw} = 15.5^m$ formed the basis for the
redshift survey of the Center for Astrophysics. Systematical measurements of
redshifts were started in 1978. By the present time, 14632 spectra have been
obtained within the frame of the CfA2 project. As to redshifts, the survey
(96\,\% completeness with $m_{\rm Zw} = 15.5^m$) comprises 13150 galaxies
with measured radial velocities.
About 30\,\% of data were taken by the authors from the literature. In the regions $20^h < \alpha < 4^h$~(south),
$8^h < \alpha < 17^h$~ (north), $-2.5^\circ < \delta < 50^\circ$ (12925
galaxies) the catalogue completeness reaches 98\,\%.

{\bf SSRS2} (July, 1998, da Costa et al., 1998). 5369 galaxies to the
apparent magnitude $15.5^m$, which cover 1.7 sterad of the southern
hemisphere, are picked up from the list of nonstellar objects catalogued in
HST\,GSC. The accuracy of determination of the positions is $\sim 1\arcsec$.
The error of photometric magnitudes is $0.3^m$. The system of magnitudes is homogeneous
over the sky and corresponds to magnitudes measured at an isophote level of
$\sim26^{m}\sq''$. The radial velocities are accurate to $\sim 40$\,km/s.
The survey completeness in redshifts is 99\,\% to the $15.5^m$ apparent magnitude.

Since SSRS2 is in fact the extension of the CfA2 survey to the southern sky,
we united in the analysis the adjacent southern and northern
regions of CfA2 and SSRS2 to obtain the following samples:
\begin{description}
\item{north:}\,
CfA2 ($0^\circ < \delta < 90^\circ$, $7^h < \alpha < 18^h$,
      $b^{II} > 15^\circ$)
+
SSRS2 (  $\delta < 0^\circ$, $b^{II} > 35^\circ$),
\item{south:}\,
CfA2($-2.5^\circ < \delta < 50^\circ$, $20^h < \alpha < 4^h$,
     $b^{II} < -15^\circ$)
+
SSRS2( $-60^\circ < \delta < -2.5^\circ$,  $b^{II} < -40^\circ$).
\end{description}

The results of the Gamma-analysis of four subsamples for the northern
and southern sky are presented in Figs. 4, 5, 6 and 7. From the independent
northern and southern samples one can see a pronounced break at 20--30 Mpc.
The large extent of the region after the break suggests that the break is due
to the distribution properties, but not the boundary effects. It should be
noted that the results of the Gamma in the northern and southern parts
are practically coincident, though the morphological distinctions (the presence
of isolated structures) are different for all samples.

\begin{figure}
\centerline{\psfig{figure=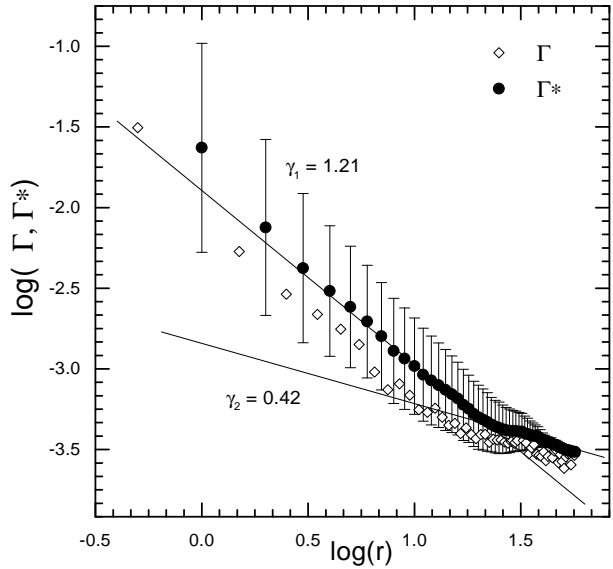,width=8cm}}
\caption{CfA2+SSRS2 north $R_{\mbox{lim}}= 140$~Mpc.}
%\label{Fig4}
\end{figure}

\begin{figure}
\centerline{\psfig{figure=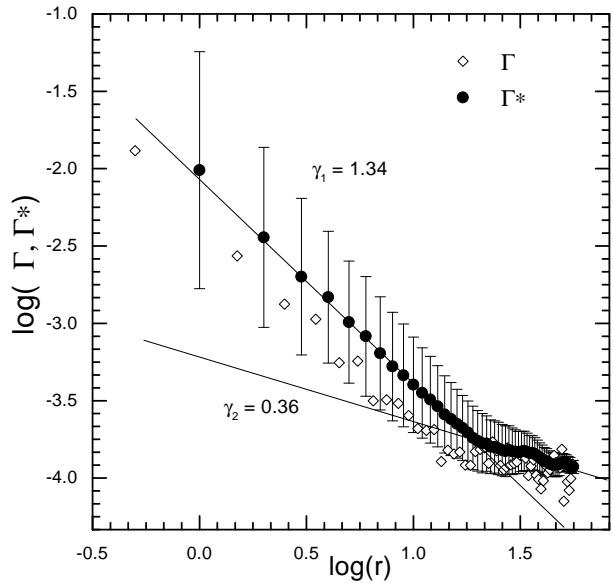,width=8cm}}
\caption{CfA2+SSRS2 north $R_{\mbox{lim}}= 160$~Mpc.}
%\label{Fig5}
\end{figure}

\begin{figure}
\centerline{\psfig{figure=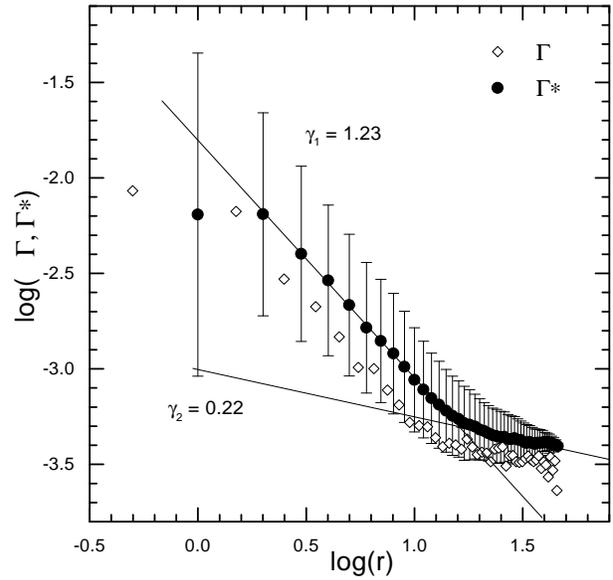,width=8cm}}
\caption{CfA2+SSRS2 south $R_{\mbox{lim}}= 140$~Mpc.}
%\label{Fig6}
\end{figure}

\begin{figure}
\centerline{\psfig{figure=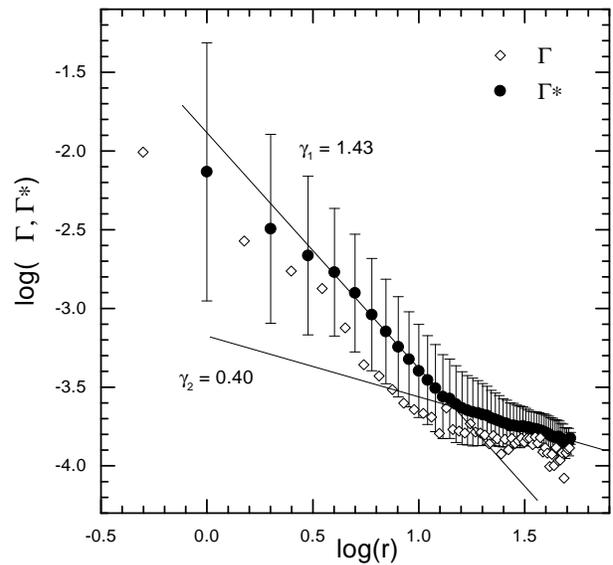,width=8cm}}
\caption{CfA2+SSRS2 south $R_{\mbox{lim}}= 160$~Mpc.}
%\label{Fig7}
\end{figure}

\subsection{APM clusters}
The catalogue of APM clusters (Dalton et al., 1992, 1997) is the first
sufficiently large catalogue of clusters of galaxies formed with the aid
of an automated procedure of isolation of objects based on the APM catalogue of galaxies to
b$_{\rm j}=20.5^m$ in the southern sky region: $-72.5\degr < \delta < -17.5\degr$,
$20.5^h < \alpha < 5.6^h$. The richness of clusters and the characteristic
brightness of galaxies were determined together in a circle of radius
$0.75~h^{-1}$\,Mpc, which is twice as small as the one used by Abell. In the
opinion of the authors of the catalogue this has improved considerably its homogeneity.
A total of 957 clusters with estimates of $z < 0.12$ are found
in this region (Dalton et al., 1997). For 374 of them the redshifts are
determined spectroscopically, including 55 clusters with measured $z > 0.12$.
The completeness of measurements of $z$ for richer clusters is considerably
higher than for poor. For this reason, we have chosen for the analysis
the APM clusters with the richness $\mathcal{R} \ge 54$, which corresponds
approximately to the richness $R\ge 0$ (or $N_A \ge 30$)  for Abell clusters.
In total such clusters number 346. Among them 43 have measured $z > 0.12$,
and we have not considered them, for 217 $z<0.12$, and for 86 there are available
only estimated $z$ (by definition smaller than 0.12). The completeness of
measurements of redshifts for the  subsample from the APM catalogue is
about 72\%.

According to the way of compilation, the clusters have been
catalogued with
the highest
statistical completeness in a certain middle interval of $z$.
A small number of nearby clusters might not be
included in the catalogue because of large angular dimensions, and a
considerable number, increasing towards the far boundary of the volume,
of distant clusters because of the decreasing with distance contrast of
the cluster and errors in the determination of the characteristic stellar
magnitude of galaxies of the cluster.

As can be seen from Figs.\,8 and 9, the Gamma shows a particularly
pronounced break at R$\simeq$32\,Mpc. The slope  $\gamma_2 \approx 0.3$,
i.e. we see a correlation of the distribution on scales $>$30\,Mpc,
but the clustering is insignificant here. The interval from the point of
break top to the limit of calculation of the Gamma is considerable, from
30\,Mpc to $\sim 100$\,Mpc. The incompleteness of the measured $z$ does not
affect the location of the break; it is only the slope of
 $\Gamma^*$ that changes prior to the break point.

\begin{figure}
\centerline{\psfig{figure=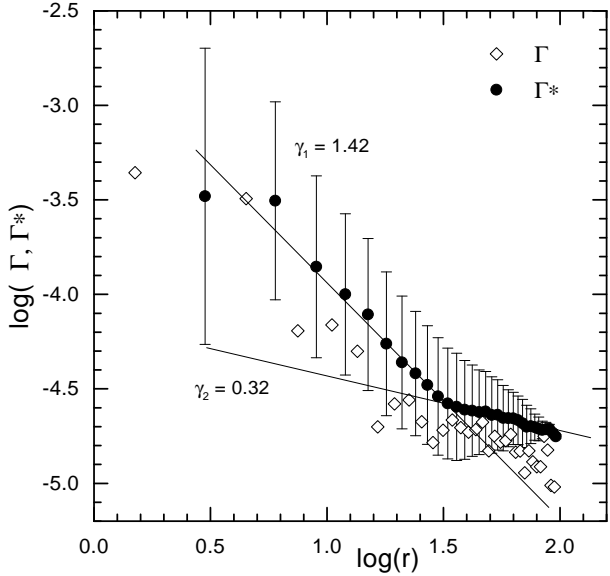,width=8cm}}
\caption{APM clusters with measured  z, $N=217$.}
%\label{Fig8}
\end{figure}

\begin{figure}
\centerline{\psfig{figure=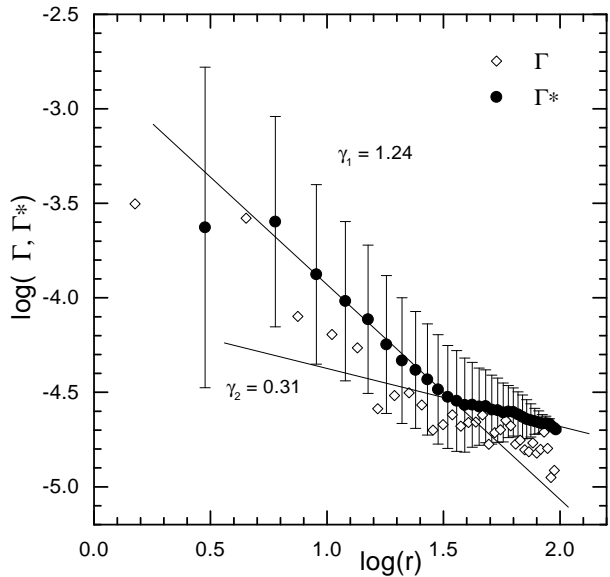,width=8cm}}
\caption{APM clusters with measured (217) and estimated (86) z, $N=303$.}
%\label{Fig9}
\end{figure}

\subsection{Rich Abell clusters in the ``Northern Cone''}
In this sample are included all  Abell's clusters of galaxies (Abell et
al., 1989) with richness
$N_A \ge 70$ ($N_A = 50-79$ corresponds to the richness class $R = 1$)
which are located in the region $b^{II} > 40^\circ$, for which $z \le 0.24$ either
measured or estimated according to Leir and van den Bergh (1977).
For 247 (77\%) out of 321 clusters there are available spectroscopically
measured
$z$, about 33\% measurements of $z$ were made with the
6\,m telescope in the course of  the programme on
studying distribution of rich clusters in the  ``Northern Cone''
(Kopylov et al., 1988; Kopylov, 1999). For such rich clusters
(remind that the clusters with $N_A \ge 80$ belong to the classes of richness
$R \ge 2$)
at high galactic latitudes ($b^{II} > 40^\circ$) the
catalogue of Abell, perhaps, may not be considered as strongly distorted
by the effects of incompleteness which could influence seriously the
shape of the Gamma. At the present time, this sample of clusters is
likely to be the best one for our purpose both in its completeness
and homogeneity and in coverage of the volume of space.

The subsample of rich clusters that we have obtained allows one to
follow the behaviour of the Gamma in a very wide and probably in the most
interesting range of scales, 10--250\,Mpc, in which from theoretical
reasoning and from the available data an asymptotic transition to the
homogeneous distribution of objects in space must be observed. It is exactly this
behaviour of the Gamma that has been detected (see Figs.10 and 11). After
the region of power law drop in density with the slope $\gamma\approx 1.5$,
starting from $\approx 40$\,Mpc, a transitional interval follows
with gradual decrease in slope, and beginning with the radius
$\geq 120$~Mpc, the distribution of clusters does not practically
differ from homogeneous up to a limiting radius of 250\,Mpc.

\begin{figure}
\centerline{\psfig{figure=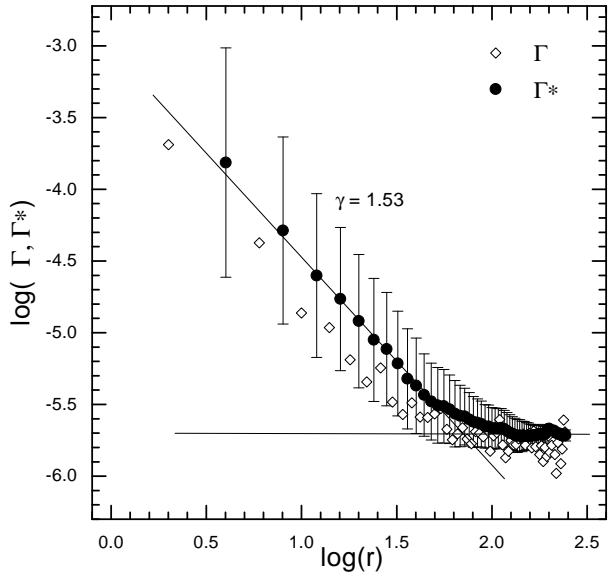,width=8cm}}
\caption{Rich Abell clusters with measured z, $N=247$.}
%\label{Fig10}
\end{figure}

\begin{figure}
\centerline{\psfig{figure=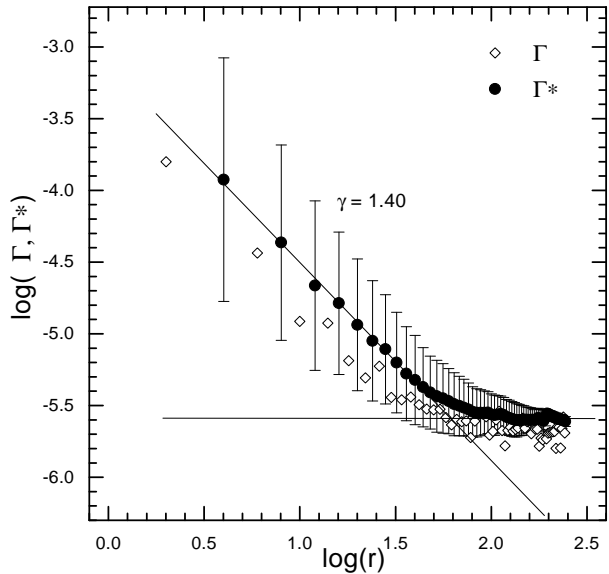,width=8cm}}
\caption{Rich Abell clusters with measured (247)
	 and estimated (74) z, $N=321$.}
%\label{Fig11}
\end{figure}

The use of the sample in which only measured $z$ are present (incomplete
sample) does not change essentially the Gamma shape and shows that the
result from the complete (in the sense of location of objects in the volume) sample
 is not practically distorted by the fact that for part of the clusters estimated
$z$ have been taken. The inclusion into the sample of clusters with estimated
$z$ leads only to reduction of the steepness of the slope on small scales,
that is, ``blurring'' of clustering of clusters on the scales of superclusters occurs.

\subsection{Superclusters of galaxies}
The catalogue of superclusters of Einasto et al. (1997) which covers the
region $R < 350$~Mpc, $|b^{II}|>17^\circ$ was constructed by the
method of percolation using  Abell clusters of richness $R\ge0$
both with measured and estimated redshifts $z < 0.12$. The percolation
radius is 24\,Mpc. A total of 220 superclusters were revealed. Einasto
et al. (1997) pointed out that superclusters are located at the
nodes of a quasi-regular lattice.

As distinct from clusters of galaxies, superclusters represent
nonvirialized systems numbering from 2 to 36 (for the richest of all ---
the Shapley supercluster) Abell clusters. The great difference in
population and, therefore, in sizes is due to the procedure of construction
of the catalogue. This may, in principle, introduce
unpredictable variations in the correlation properties of the sample
since the superclusters considered in the calculations as point objects
may actually be strongly anisotropic systems, for revealing of which the
method of percolation is just the most suitable. Nevertheless, the Gamma
calculation for the superclusters of Einasto et al. is of interest
as an independent method of control of clustering of matter on
sufficiently large scales ($\leq 150$\,Mpc).

\begin{figure}
\centerline{\psfig{figure=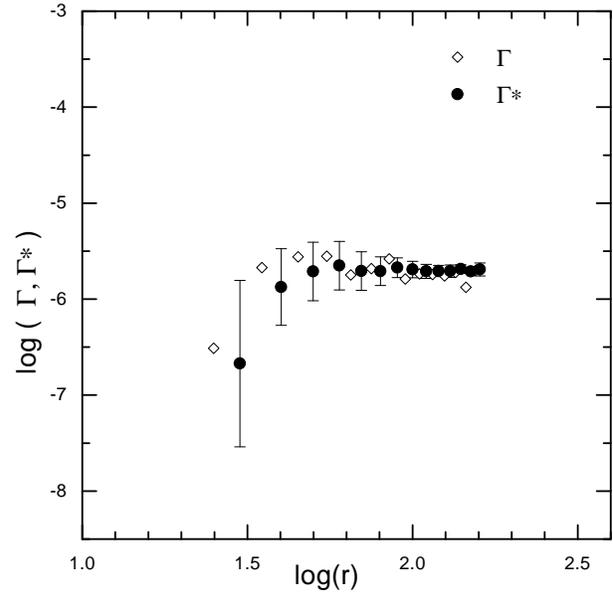,width=8cm}}
\caption{Einasto superclusters, north.}
%\label{Fig12}
\end{figure}

\begin{figure}
\centerline{\psfig{figure=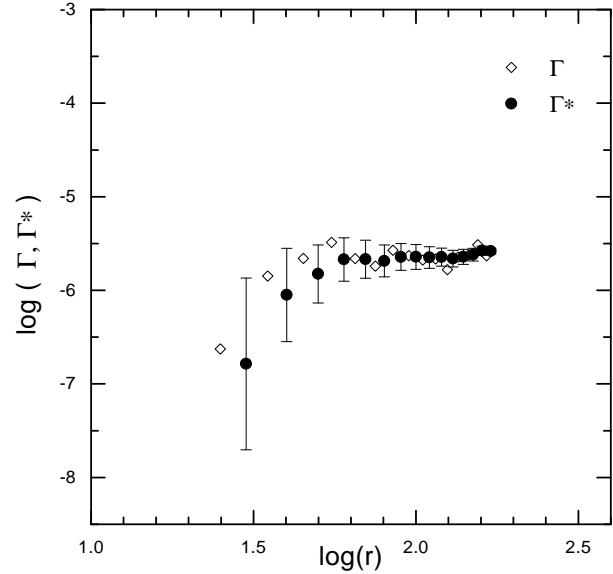,width=8cm}}
\caption{Einasto supercluster, south.}
%\label{Fig13}
\end{figure}

\begin{table*}
\begin{center}
\caption{Parameters of various samples}
%\label{Tab1}}
\begin{tabular}{ccccccc}
\hline
$R_{\mbox{lim}}$&
$M_{\mbox{lim}}$ &
N &
$R_{\mbox{s}}$&
$\gamma_1$ &
$R_{\mbox{break}}$&
$\gamma_2$ \\
(Mpc)& & & (Mpc)& & (Mpc)& \\
\hline
\multicolumn{7}{l}{Local volume ($H_0=70$)}\\
        7.15  & -13.0 & 198 &   7.15 & 1.19 & -- &  --  \\
\multicolumn{7}{l}{Local ``disk'' ($H_0=70$)}\\
        2.0   & -13.0 & 146 &   2.0  & 1.44 & -- &  --  \\
\hline
\multicolumn{7}{l}{CfA2+SSRS2, north ($H_0=100$)}\\
        180&    -20.78& 212     &69&    1.41&   24&     0.52\\
        160&    -20.52& 458     &69&    1.34&   20&     0.36\\
        140&    -20.23& 959     &61&    1.21&   23&     0.42\\
        120&    -19.90& 1691    &52&    0.98&   19&     0.46\\
        100&    -19.50& 2610    &43&    0.86&   16&     0.40\\
\multicolumn{7}{l}{CfA2+SSRS2, south ($H_0=100$)}\\
        180&    -20.78& 194     &64&    1.53&   25&     0.11\\
        160&    -20.52& 339     &58&    1.43&   15&     0.40\\
        140&    -20.23& 611     &52&    1.23&   17&     0.22\\
        120&    -19.90& 1056    &44&    0.97&   20&     0.19\\
        100&    -19.50& 1517    &39&    1.01&   19&     0.19\\
\hline
\multicolumn{7}{l}{APM-clusters, richness $\mbox{$\mathcal{R}$} \ge 54$ ($H_0=100$)}\\
\multicolumn{7}{l}{ 1)with measured and evaluated $z$ }\\
        339&          & 303&    105&    1.24&   32&     0.31\\
\multicolumn{7}{l}{ 2) only with measured $z$ }\\
        339&          & 217&    105&    1.42&   32&     0.32\\
\hline
\multicolumn{7}{l}{Abell clusters, richness $N_A \ge 70$ ($H_0=100$)}\\
\multicolumn{7}{l}{ 1) with measured and evaluated $z$ }\\
        638&          & 321&    246&    1.40&$\approx 40$&  $\sim 0.3$\\
           &          &    &       &        &$\ge 120$& $\simeq 0$\\
\multicolumn{7}{l}{ 2) only with measured $z$ }\\
        638&          & 247&    246&    1.53&$\approx 40$&  $\sim 0.3$\\
           &          &    &       &        &$\ge 120$& $\simeq 0$\\
\hline
\multicolumn{7}{l}{Einasto superclusters, north ($b^{II} > 15^\circ$) ($H_0=100$)}\\
        350&          &  98&    164&      --&     (40)&   ~0 \\
\multicolumn{7}{l}{Einasto superclusters, south ($b^{II} < 15^\circ$) ($H_0=100$)}\\
        350&          & 122&    173&      --&     (40)&   ~0 \\
\hline
\end{tabular}
\end{center}
\end{table*}

In Figs.\,12 and 13 is displayed the Gamma calculated for the northern
and southern subsamples of superclusters. On scales larger than the
percolation radius, immediately after the appearance of the ``signal'',
that is when a sufficient number of objects fall within the sphere
and spherical layers, the Gamma shows directly  homogeneity of
distribution of superclusters in both the southern and northern
regions beginning with a scale of about 40\,Mpc, that is, approximately
from the scale on which the Gamma of clusters of galaxies shows the
first break. The second, a less pronounced break (or a feature on the
curve of the Gamma variation) on a scale of about 120\,Mpc may likely
be associated with transition to quasi-regular distribution of the superclusters
on a lattice with a period of about 125\,Mpc (Einasto et al., 1997).
Note that evidence to this scale of clustering in the
spatial distribution of rich clusters was also obtained in measuring the
correlation function of rich compact clusters in the ``Northern Cone''
(Kopylov et al., 1988).

\section{Discussion of results}
In Table 1 are collected the principal results of the analysis. The
first four columns present the basic parameters of the samples of objects.
$R_{\mbox{lim}}$ is the far boundary of the sample along the radial
coordinate. $M_{\mbox{lim}}$ is the limitation on absolute magnitude for
the ``volume limited'' sample, which corresponds to $R_{\mbox{lim}}$.
$N$ is the number of objects in the sample. $R_s$ is the radius of the
maximum sphere falling inside the boundaries of the sample. In the last
three columns are given the parameters of the simplest model that we
have used to describe the Gamma behaviour: ``power law drop'' of density
(index $\gamma_1$) --- break ($R_{\mbox{break}}$) --- change
of the mode of clustering ($\gamma_2$)''.

\begin{figure*}
\centerline{\psfig{figure=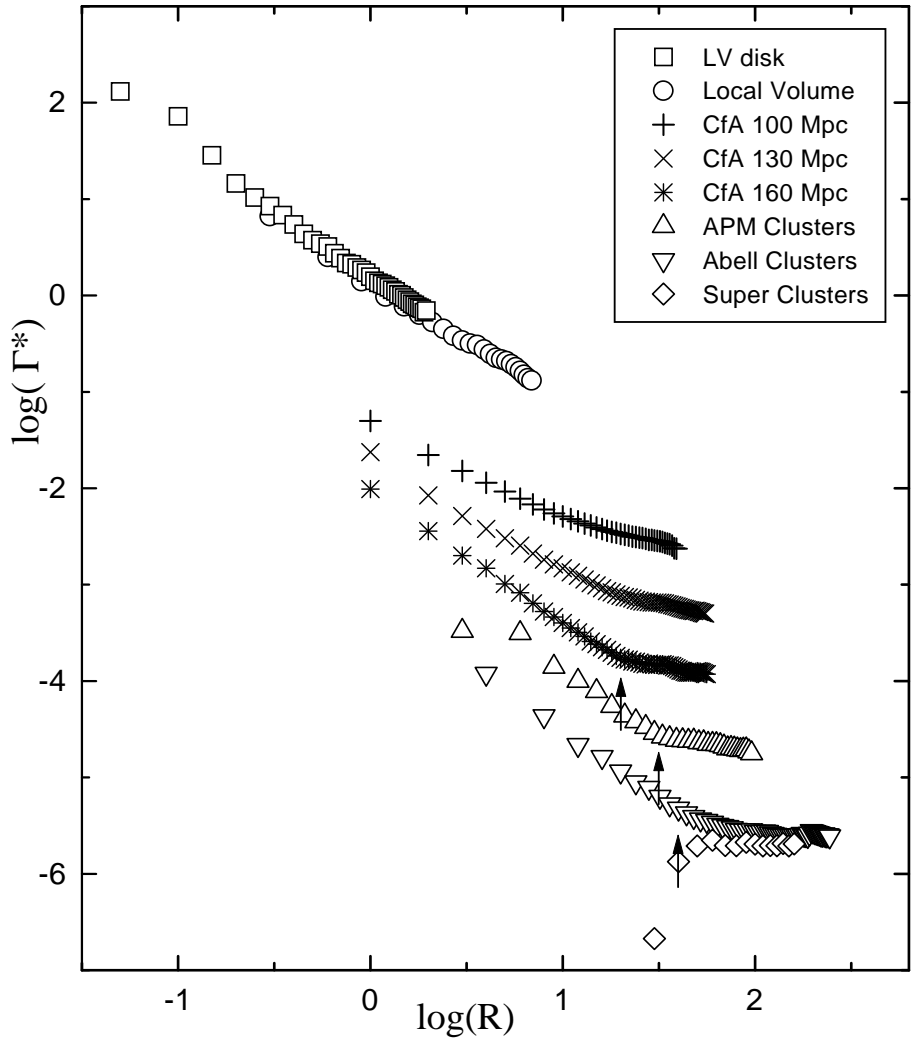,width=12cm}}
\caption{A master figure of Gamma functions for different samples.
 Squares and circles show the behaviour of the Gamma for samples
of the Local Volume and Supergalactic Disk, respectively.
The crosses of different form  are for the northern part of the samples
 CfA2+SSRS2, for the depths
  100, 130, 160~Mpc. The triangles are for APM clusters. The inverted
triangles are for Abell clusters. The diamonds designate the Gamma for
Einasto superclusters.}
\label{Fig14}
\end{figure*}

The results of investigation of all principal samples are displayed in
Fig.\,14 in order to make more convenient their comparison with one
another and to attempt to reveal common laws of clustering throughout
the investigated range of scales.

In Fig.\,14 one can see an approximately ``coordinated'' break over
different samples including the spheres with $R > 30-40$\,Mpc
(the first arrow). From the galaxies of the northern part of the sample
CfA2+SSRS2 the extension of the region of the Gamma after the break
suggests that the break is most likely caused not by the boundary
effects, but rather by the properties of distribution. For APM
clusters the Gamma has a break at $\approx 30$\,Mpc, the limiting
scale is 105\,Mpc. The tendency of behaviour of this type Gamma
(break and transition to other mode
of clustering) is terminated  by deep samples
of rich clusters and superclusters, nearly complete ``leveling out''
of the Gamma (the second and third arrows) is observed. A trend of
slope $\gamma_1$ (prior the break) is noticeable over all
samples, the dynamics and causes of which still remain to be solved.
Formally, the power law of the drop of density begins with scales
 $\sim50$\,kpc, immediately after the size of a huge galaxy.

We used rather heterogeneous data, even though the results obtained
from different samples basically agree with each other in the main
features, which allows some general conclusions to be drawn concerning
the character of behaviour of the Gamma on different scales.
\begin{enumerate}
\item The shape of the correlation Gamma function  points to the
density variation according to a power law.
\item The power index on small scales (prior the break) changes
within 0.9--1.5. In principle, this corresponds to  fractal distribution,
but the scatter in  slopes for different samples of objects of
different nature makes questionable the possibility of describing a
large-scale structure by a single fractal dimension.
\item Manifestations of fractality have been revealed on scales spanning
nearly 3 orders of magnitude --- from 50\,kpc to 30\,Mpc.
\item Systematic variations of the Gamma parameters obtained from
CfA2+SSRS2 galaxies require additional investigation, although the tendency to
``leveling out'' of the Gamma at $R > 20-25$\,Mpc for brighter objects
is consistent with the results for clusters (brighter galaxies
concentrate towards the centres of clusters).
\item On scales  30--40\,Mpc, in the cases where the Gamma is
calculated to  scales well larger than 40\,Mpc, a  break is seen
in the behaviour of the Gamma (index $\gamma \leq 0.5$) which implies
the change of the condition and, possibly, a physical mechanism of
clustering.
\item From 30 to 100--120\,Mpc the clustering is likely to occur,
since on these scales isolated structures (Great Wall, voids, etc.)
are observed, but it is insignificant, for when increasing the scales,
the density drops slightly. Probably, the contrast (amplitude) of
inhomogeneities on these scales is already small.
\item An asymptotic transition to homogeneity $\gamma_2\rightarrow0$ might be
supposed, mainly from the distribution of rich clusters and superclusters
on scales  40--120\,Mpc.
\end{enumerate}

As a whole, the correlation Gamma-function is quite an interesting and
informative way of describing  a large-scale structure. It should be
noted, however, that the influence of selection effects on the shape of
the Gamma is still not clearly studied. A study is required of model
samples with expected properties of the Gamma and properties introduced
by real selection effects (for instance, the density gradient along
the radial coordinate).

The limitation of scales of calculation by spheres located inside the
sample boundaries does not solve the problems of boundary conditions
completely, because the Gamma becomes dependent on the location of
significant structures inside the sample boundaries. This points to
the necessity of careful selection of objects and choice of sample
boundaries. The incompleteness of the sample may  diminish
the statistical significance of the result of Gamma application.
Other methods should be involved to confirm the interpretation.

Nevertheless, it can be argued with assurance that our results
of application of the Gamma function are in fair agreement with
theoretical considerations of the finitness of the range of scales
with the fractal type of correlation of density and of gradual
decrease of contrast of inhomogeneities with increasing volume
of the region of the Universe being studied.
\begin{acknowledgements}
The authors express their gratitude to Yu.V.\,Baryshev, V.A.\,Hagen-Thorn
and I.D.\,Karachentsev for keen interest they took in the work and
assistance in its accomplishment. A.Tikhonov is grateful for support through
the grant
``Integration'' N A0007.
\end{acknowledgements}

\end{document}